\documentclass[fleqn,usenatbib]{mnras}

\usepackage{newtxtext,newtxmath}

\usepackage[T1]{fontenc}
\usepackage{ae,aecompl}

\usepackage{graphicx}	
\usepackage{amsmath}	
\usepackage{gensymb} 	
\usepackage{float}      
\usepackage{orcidlink}  
\usepackage{comment}    

\def\vel{\mathrm{v}}                   
\def\AN{\mathbf{\Omega}}               
\newcommand\vhat[1]{\overset{\rightharpoonup}{#1}}

\title[Grid-Based Polar Circumbinary Disks]{Grid-Based Simulations of Polar Circumbinary Disks: Polar Alignment and Vortex Formation}

\author[I. Rabago et al]
{
Ian Rabago$^{1,2}$ \orcidlink{0000-0001-5008-2794},
Zhaohuan Zhu$^{1,2}$ \orcidlink{0000-0003-3616-6822},
Rebecca G. Martin$^{1,2}$ \orcidlink{0000-0003-2401-7168} and
Stephen H. Lubow$^{3}$ \orcidlink{0000-0002-4636-7348}
\\
$^{1}$Department of Physics and Astronomy, University of Nevada, Las Vegas 4505 S. Maryland Parkway Las Vegas, NV 89154, USA\\
$^{2}$Nevada Center for Astrophysics, University of Nevada, Las Vegas 4505 S. Maryland Parkway Las Vegas, NV 89154, USA\\
$^{3}$Space Telescope Science Institute, Baltimore, MD 21218\\
}

\date{Accepted XXX. Received YYY; in original form ZZZ}

\pubyear{20XX}

\begin{document}
\label{firstpage}
\pagerange{\pageref{firstpage}--\pageref{lastpage}}
\maketitle

\begin{abstract}
We describe the first grid-based simulations of the polar alignment of a circumbinary disk.  We simulate the evolution of an inclined disk around an eccentric binary using the grid-based code ATHENA++.  The use of a grid-based numerical code allows us to explore lower disk viscosities than have been examined in previous studies.  We find that the disk aligns to a polar orientation when the $\alpha$ viscosity is high, while disks with lower viscosity nodally precess with little alignment over 1000 binary orbital periods.  The timescales for polar alignment and disk precession are compared as a function of disk viscosity, and are found to be in agreement with previous studies.  At very low disk viscosities (e.g. $\alpha = 10^{-5}$), anticyclonic vortices are observed along the inner edge of the disk.  These vortices can persist for thousands of binary orbits, creating azimuthally localized overdensities as well as multiple pairs of spiral arms.  The vortex is formed at $\sim 3-4$ times the binary semi-major axis, close to the inner edge of the disk, and orbits at roughly the local Keplerian speed.  The presence of a vortex in the disk may play an important role in the evolution of circumbinary systems, such as driving episodic accretion and accelerating the formation of polar circumbinary planets.
\end{abstract}


\section{Introduction}
\label{sec:introduction}
Protoplanetary disks are produced during the star formation process.  Stars are frequently formed in pairs or clusters during the collapse of a molecular cloud \citep{Duquennoy1991,Kouwenhoven2007}, and so disks around multiple stars are expected to be common \citep{Monin2007}.  Analogous to how circumstellar protoplanetary disks can give rise to planets, the circumstellar and circumbinary disks in a binary star system may act as the birthsites of the S-type and P-type circumbinary planets, respectively \citep{Dvorak1986}.  Observations of binary star systems at different evolutionary stages have revealed different types of protoplanetary disks and planetary systems, including circumbinary disks \citep{Simon1992,Czekala2017,Kennedy2019,Bi2020,Kraus2020}, individual circumstellar disks \citep{Cruz2019,Keppler2020}, and exoplanet systems in both S- and P-type orbits (e.g. \citealt{Doyle2011,MartinD2018}).

Binary and higher-order star systems exhibit rich dynamics that are not seen in single-star systems.  The gravitational torque from the inner binary creates a region of dynamical instability, limiting the range of stable orbits for S- and P-type planets \citep{Holman1999,Quarles2018,Chen2020}.  Companion stars on faraway orbits can perturb the system over long timescales, causing orbiting particles to undergo von Zeipel-Lidov-Kozai oscillations \citep{vonZeipel1910,Kozai1962,Lidov1962}.  For simplicity, we restrict most of the discussion in this paper to the binary case.

For cicumbinary or P-type orbits around a circular orbit binary, misalignment of the orbit to the binary orbital plane  causes the orbit to nodally precess about the binary angular momentum vector.  If the inner binary is eccentric, a second stable configuration exists in which the orbiting body precesses about the binary eccentricity vector in a polar orientation \citep{Verrier2009,Farago2010,Doolin2011}.  For circumbinary disks around the eccentric binary, inclination damping due to viscous forces  causes the disk to settle at either $0\degree$ or $90\degree$, depending on the initial misalignment.  This can lead to the creation of polar disks, where the disk is aligned perpendicular to the orbits of the binary stars.  The stability of these two orientations has previously been confirmed using analytic theory and SPH numerical simulations \citep{Martin2017,Martin2018,Lubow2018,Zanazzi2018,Smallwood2019}.

A handful of such ``polar disks'' have been observed. The HD 98800 system is a hierarchical double binary system with a circumbinary disk observed around the B binary component \citep{Kennedy2019}.  The A and B binary systems orbit each other on a $67$ AU orbit with moderate eccentricity $(e \sim 0.5)$, while the disk-hosting BaBb binary components orbit on a $1$ AU orbit with high eccentricity $(e \sim 0.8)$ \citep{Zuniga2021}.  The HD 98800 B disk is truncated on both edges by the inner and outer binary, with the dust disk extending between 2.5 and 4.6 AU \citep{Kennedy2019}. More recently, a polar circumbinary disk was observed around V773 Tau B \citep{Kenworthy2022}.  Observations of the IRS 43 system \citep{Brinch2016} have found an edge-on circumbinary disk with the binary star components orbiting outside of the circumbinary disk plane.  Both components of the binary are also surrounded by circumstellar disks, each at a different angle to the circumbinary disk, suggesting complex accretion from the circumbinary disk onto the central stars.

To date, all simulations of polar disks have been performed using smoothed-particle hydrodynamics (SPH)  (e.g., \citealt{Martin2017, Kennedy2019, Smallwood2019,Cuello2019,Martin2022}).  These simulations typically make use of over $10^6$ particles and are able to capture the global dynamical evolution of the disk.  However, the range of viscosities that can be included in SPH simulations is limited by the number of particles in a given region. The minimum disk effective viscosity \citep{Shakura1973} is roughly $\alpha = 0.01$ for most simulations \citep{Price2018}.  Protoplanetary disks around single stars have been found with viscosities of $\alpha  \lesssim 10^{-4}$ \citep{Pinte2016,Villenave2022}, and so the study of how polar disks behave at lower viscosities is important but has been unexplored.

In this paper we work to expand on previous results using a grid-based hydrodynamic code to simulate the circumbinary disk.  This allows us to extend the results of previous works to lower viscosities than have previously been examined using SPH simulations.  Our paper is organized as follows.  In Section~\ref{sec:methods}, we outline our computational setup.  Our results are presented in Section~\ref{sec:results}.  We discuss the implications to circumbinary disk evolution in Section~\ref{sec:discussion}, and conclude in Section~\ref{sec:conclusion}.

\section{Methods}
\label{sec:methods}

\begin{figure}
    \centering
    \includegraphics[width=\columnwidth]{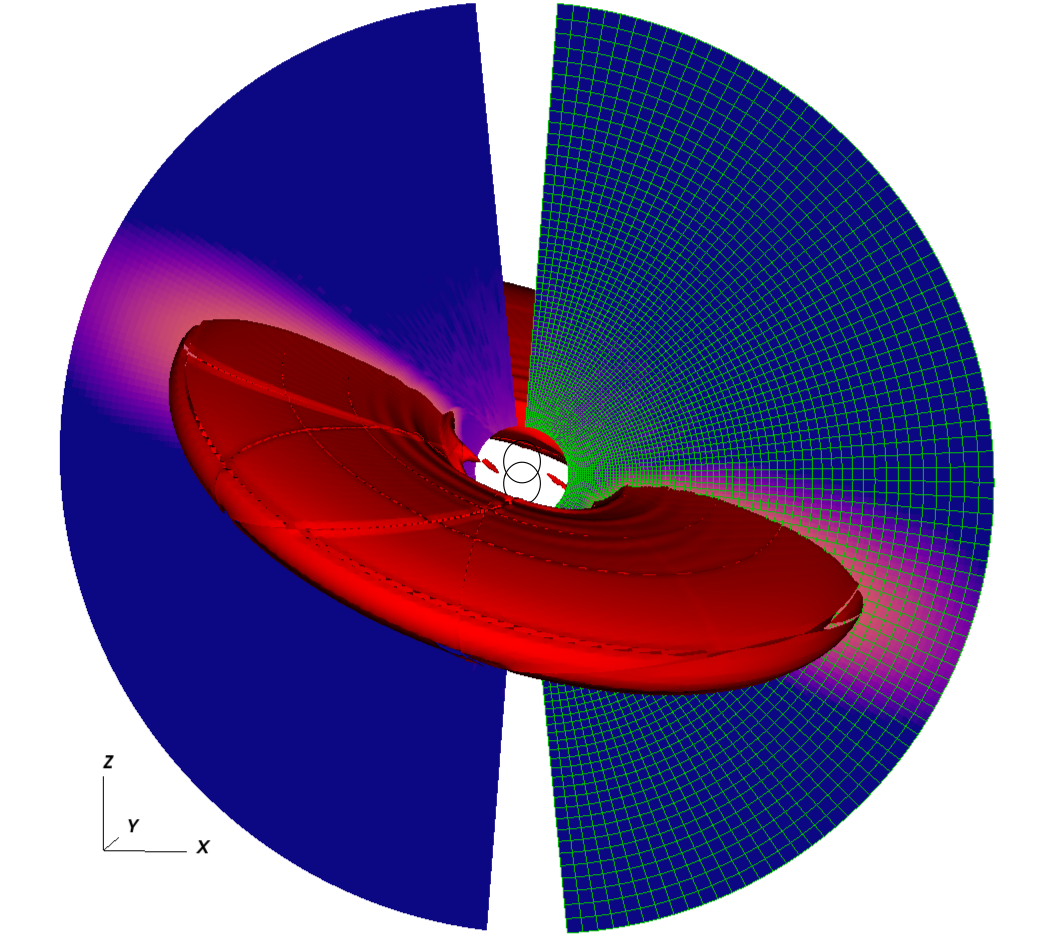}
    \caption{ Visualization of the $\alpha = 10^{-2}$ simulation setup at $t = 1000$ binary orbits.  The disk starts inclined at $t = 0$ by rotating the disk along the x-axis, keeping $\vhat{L}_{\rm disk}$ in the yz-plane, and undergoes nodal precession about the z-axis due to the binary torque.  Colors show a slice of density along the xz-plane, with red contour showing a 3D isosurface.  The orbits of the binary stars are visible in the center.  The simulation grid is shown at half resolution on the right half of the simulation domain.  Regions close to the binary and the z-axis are not in the computational domain.  An animation of the disk evolution is available in the online material associated with this paper and can be downloaded at \url{https://youtu.be/Ny-ggFHALMA}. }
    \label{fig:setup}
\end{figure}

\begin{figure}
    \centering
    \includegraphics[width=\columnwidth]{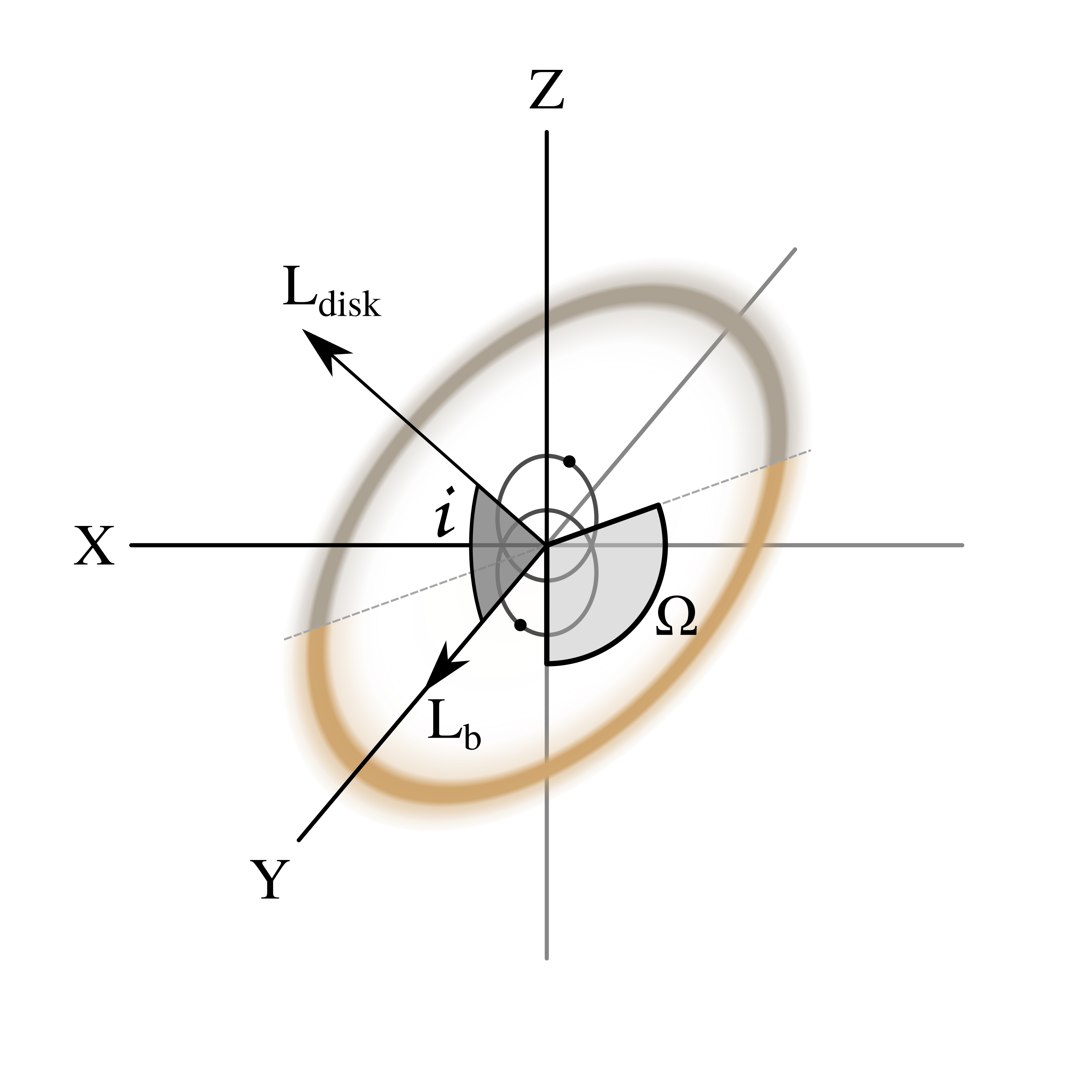}
    \caption{Schematic of the central binary and circumbinary disk, as well as the angles used to measure the disk orientation.  The binary orbits in the $xz$-plane, with the angular momentum vector $\vhat{L}_{\rm b}$ pointing along the positive y-axis and the eccentricity vector $\vhat{e}_{\rm b}$ pointing along the z-axis.  The disk is oriented in 3D space by its angular momentum vector $\vhat{L}_{\rm disk}$, where it forms an angle $i$ with the binary angular momentum vector $\vhat{L}_{\rm b}$.  The gray dashed line denotes the disk's ``line of nodes'', where the disk crosses the $xz$-plane at an angle $\AN$ with the eccentricity vector $\vhat{e}_{\rm b}$.}
    \label{fig:geometry}
\end{figure}

We simulate a circumbinary disk using the grid-based hydrodynamical code \textsc{ATHENA}++ \citep{Stone2020}, using spherical-polar coordinates $(r, \theta, \phi)$ for the simulations.  Figure \ref{fig:setup} shows a visualization of our simulation setup.  A disk that begins with an inclination in the range for polar alignment evolution nodally precesses about the binary eccentricity vector.  Therefore, we place the binary in the $xz$-plane of the simulation, so that the binary angular momentum vector $\vhat{L}_{\rm b}$ points along the positive y-axis and the eccentricity vector $\vhat{e}_{\rm b}$ points along the positive z-axis.  With this placement of the binary, a disk aligning towards a ``polar'' orientation  settles towards the $xy$-plane of the simulation domain $(\theta = \pi/2)$.  This binary orientation keeps the disk from precessing outside of the specified domain during the simulation.  This orientation has the additional benefit of orienting the major axes of the binary towards the poles of the spherical-polar simulation domain, which prevents strong gravitational interactions from occurring close to the inner radial boundary.





The primary simulation domain spans a range of $r = [1.07a_b, 10.7a_b]$, $\theta = [5 \degree, 175 \degree]$, and $\phi = [0, 2\pi]$.  We use a grid of 144 logarithmically spaced cells in $r$, 176 uniformly spaced cells in $\theta$, and 384 uniformly spaced cells in $\phi$.  Our disk setup is based on the setup used in \cite{Martin2017}.  We initialize the midplane density profile of the disk according to the density power-law profile 

\begin{equation}
    \rho(R, z = 0) = \rho_0 \left( \frac{R}{R_0}\right)^{-p}
\end{equation}
and set the vertical profile by numerically integrating the density at each grid cell to establish hydrostatic equilibrium according to the disk scale height $H = c_s/\Omega_K$.  Here, $R$ is the cylindrical radius to the disk axis of rotation, $R_0 = a_b$, $\rho_0 = 1$ at $R_0$, $c_s$ is the local isothermal sound speed $\sqrt{P/\rho}$ at $R$, and $\Omega_K$ is the Keplerian orbital frequency at $R$.  The disk temperature is initialized using the power-law profile
\begin{equation}
    T(r) = T_0 \left( \frac{r}{r_0}\right)^{-q},
\end{equation}
where $r_0 = a_b$ and $T_0 = 1$ at $R_0$.  We use power-law exponents of $p=2.25$ and $q=1.5$, corresponding to an initial disk surface density profile $\Sigma \propto r^{-3/2}$ and a constant kinematic viscosity $\nu = \alpha c_{\rm s}^2/\Omega_K$.  The density profile is truncated outside of the range $[2a_b, 5a_b]$ using an exponential cutoff of the form $\exp{\left( - |r-\mu_r|/\sigma_r \right) }$, where $\mu_r$ are the disk edges of $2a_b$ and $5a_b$ and the relative scale length $\sigma_r$ is $\sigma_{r,\rm in} = 0.35a_b$ for the inner edge and $\sigma_{r,\rm out} = 1.8a_b$ for the outer edge.  

We initialize the disk with a spherically isothermal temperature profile, as well as an initial scale height of $H/r = 0.1$ at $R = 3.5a_b$.  We use the orbital cooling scheme outlined in Equation 5 of \cite{Zhu2015}, with a dimensionless cooling time of $t_{\rm cool} = 0.01\Omega_K$.  At the simulation boundaries, we instate a one-way outflow boundary condition in the radial direction and a reflecting boundary condition in the polar direction.  We use a spherically symmetric density floor with a value of $\rho_{\rm floor} = 10^{-4} \rho_0$ at $r = 1$ and a power-law slope of $d=2.25$, identical to that of the density profile.  
The disk velocity is initialized with a Keplerian profile using the total mass of the binary $M_{\rm tot} = m_1 + m_2$.  Since the axis of rotation is now within the simulation domain, regions near the poles of rotation are initialized with a velocity
\begin{equation}
  \vel = \sqrt{\frac{GM_{\rm tot}}{r}}\frac{R}{r},
\end{equation}
which prevents excessively high velocities occurring near the poles of rotation.


The binary components are simulated as gravitational bodies, with equal masses $m_1 = m_2 =0.5$, which we place in an eccentric orbit with semi-major axis $a_b$ and eccentricity $e_b$.  We choose an eccentricity of $e_b = 0.5$. 
This sets the initial tilt required for disk polar alignment of a low mass disc to be  
$ i_{\rm crit} < i < 180\degree - i_{\rm crit}$, where $i_{\rm crit}= \arcsin{\sqrt{3/8}} \simeq 38\degree$, based on the behavior of test particles \citep{Farago2010}.  For massive disks, the critical angle changes due to angular momentum exchange between the disk and the binary \citep{Martin2019}.  Since the binary does not feel the gravitational force from the disk and disk self-gravity is not included from our simulations, the disk can be considered to be in the low mass regime.  We use a 2nd-order leapfrog integrator to solve the equations of motion for the binary. We use the CFL fluid timestep of \textsc{ATHENA}++ as the timestep for the integrator.





Figure \ref{fig:geometry} shows the geometric setup of the binary-disk system, with inclination angle $i$ and ascending node $\AN$ labeled.  The angles $i$ and $\AN$ are measured relative to the binary plane using the equations
\begin{equation}
    i = \arccos{\left( \frac{L_y}{L} \right)}
    \label{eq:inc}
\end{equation}
and
\begin{equation}
    \AN = \arctan{\left( \frac{L_z}{L_x} \right)},
    \label{eq:omega}
\end{equation}
where $L_x$, $L_y$ and $L_z$ are the components of the disk angular momentum $L_{\rm disk}$ in the simulation $x$, $y$ and $z$ axes.

We initialize the disk at an inclination of $i = 120\degree$ and ascending node $\AN=90^\circ$ with respect to the plane of the binary by converting to tilted disk coordinates (\citealt{Zhu2019}, Eq. 44, also see Appendix \ref{sec:appendix}).  The high initial inclination ensures the disk starts within the librating region that will evolve towards a polar alignment.  For test particles around an equal-mass binary, there is no difference between the choice of prograde $(60\degree)$ versus retrograde $(120\degree)$ orbits, as the evolution of the test particle only depends on whether the angle between the binary and disk planes is beyond the critical inclination angle $i_{\rm crit}$.  
When full three-body systems are considered, angular momentum can be exchanged between the binary system and the circumbinary particle \citep{Farago2010,Martin2019}.  This creates a difference between the prograde and retrograde cases, and large angular momentum ratios $j = L_{\rm disk}/L_{\rm b}$ can lead to the formation of ``crescent'' librating orbits \citep{Chen2019,Abod2022}.  Since we do not model the exchange of angular momentum between the disk and the binary, we expect the initially prograde and retrograde cases to be similar.

We simulate the disk evolution for 1000 binary orbits, using the vl2 second-order van Leer predictor-corrector time integrator \citep{Stone2009} and third-order PPM reconstruction.  We perform a set of five simulations, varying the $\alpha$-viscosity parameter between $\alpha = 10^{-1}, 5\times10^{-2}, 10^{-2}, 10^{-3}$, and $10^{-5}$.  The disk viscosity is constant throughout most of the simulation domain.  Close to the inner radial boundary, where $R < R_{\rm{min}} = 1.07a_b$, we reduce the value of the kinematic viscosity $\nu$ using an exponential cutoff

\begin{equation}
    \nu = \alpha \frac{c_{\rm s}^2}{\Omega_K} \exp{\left( \frac{R - R_{\rm min}}{0.268} \right)}.
\end{equation}
This prevents the diffusive timestep from becoming too small and restricting the simulation speed. 
For simulations with $\alpha > 0.01$, the viscous timescale $t_{\rm visc} \sim R^2/\nu$ is short enough that material spreading of the disk requires that we expand our simulation region. For these simulations, we expand the outer radial edge of our simulation domain to $R_{\rm out} = 35.7a_b$ using 216 logarithmically spaced radial cells in order to match the increased viscous spreading of the disk.


\section{Results}
\label{sec:results}

\begin{figure}
    \centering
    \includegraphics[width=\columnwidth]{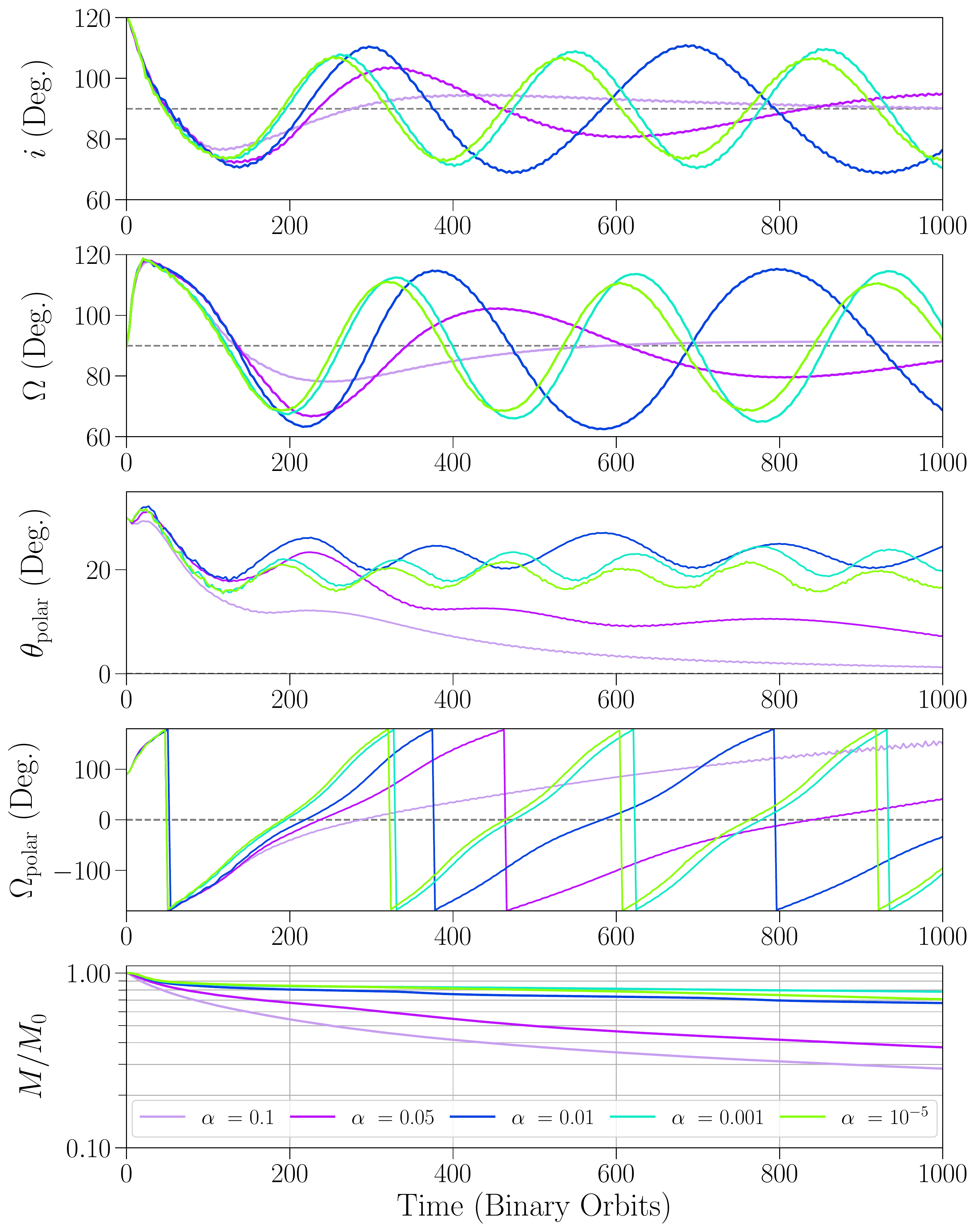}
    \caption{Time evolution of the disk for different values of $\alpha$, showing large changes in orientation.  \emph{Top:} Disk inclination $i$.  \emph{Top Middle:} Longitude of ascending node $\AN$.  \emph{Middle:}  Angle between the disk and the binary eccentricity vector $\theta_{\rm polar}$.  \emph{Bottom Middle:}  Longitude of ascending node in the xy-plane $\AN_{\rm polar}$.
    \emph{Bottom:} Disk mass as a fraction of the initial mass.  All angular quantities are measured in reference to the binary orbital plane at a distance of $R = 3a_b$. 
    }
    \label{fig:angle}
\end{figure}

\begin{figure}
    \centering
    \includegraphics[width=\columnwidth]{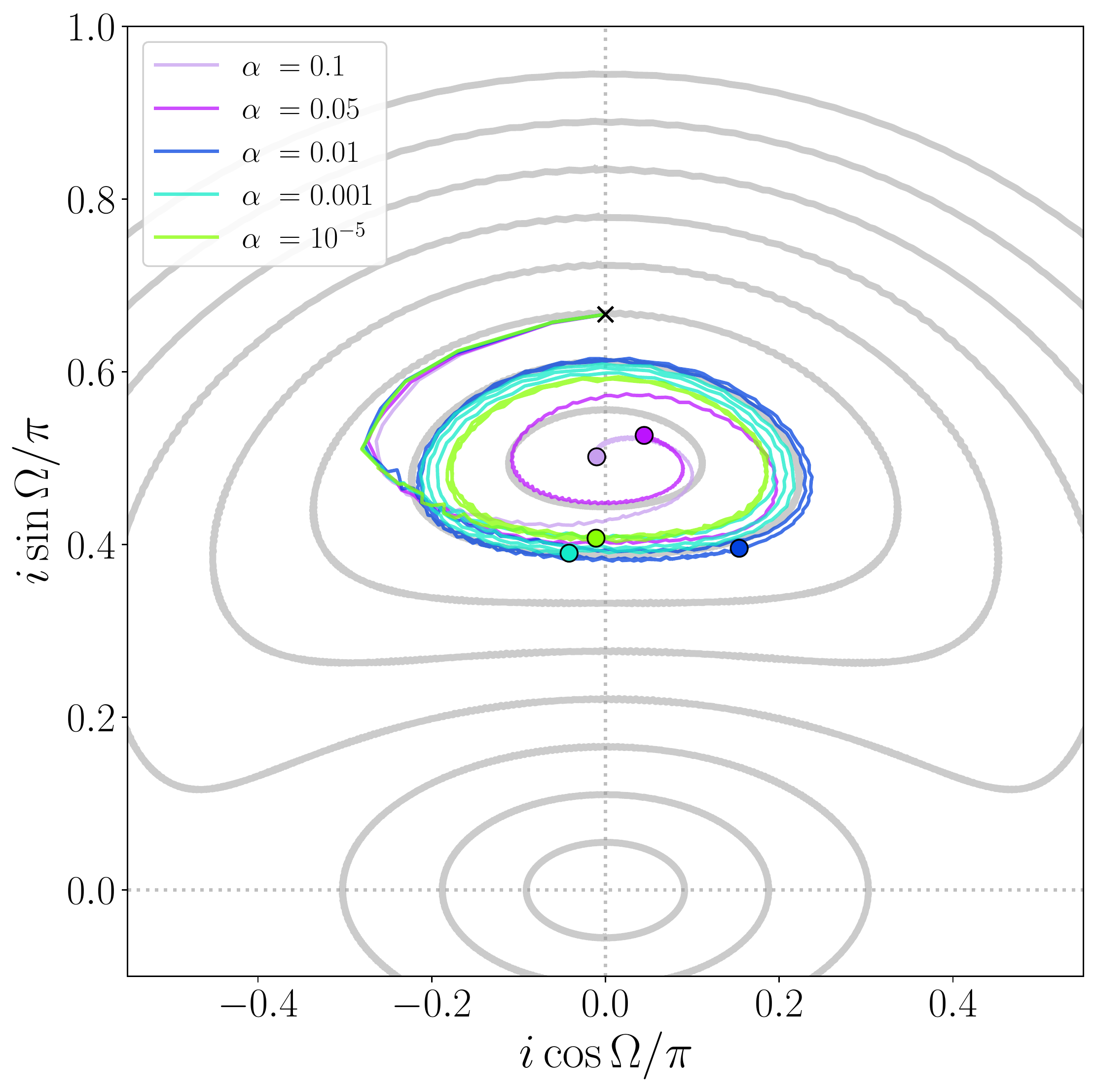}
    \caption{Trajectories of the simulations in $i\mathrm{\cos\AN} - i\mathrm{sin\AN}$ phase space.  Each trajectory starts at the black `$\times$' and spirals inwards counterclockwise towards the point $(0.0,0.5)$, with the final state of each simulation at $1000$ binary orbits marked by the colored dots. Angular quantities are measured in reference to the binary orbital plane at a distance of $R = 3a_b$, as in Figure \ref{fig:angle}.  Gray paths show the trajectories of test particle orbits, spaced by $10\degree$.  An animation of the test particle precession, as well as the projection onto the phase space, is available with the online version of this paper and can be downloaded at \url{https://youtu.be/ZBkhG6pkOAQ}.}
    \label{fig:phase}
\end{figure}

\subsection{Polar Alignment}
\label{sec:alignment}

We observe polar libration of the disk in all of our simulations.  The top two panels of Figure \ref{fig:angle} show the disk inclination and ascending node oscillating about $i=\AN=90 \degree$, with a precession time $t_{\rm prec}$ that increases with  disk viscosity.  The third and fourth panels of Figure \ref{fig:angle} show the polar libration of the disk in reference to the simulation xy-plane, using the angles $\theta_{\rm polar}$ and $\AN_{\rm polar}$, the angle between $\vhat{L}_{\rm disk}$ and $\vhat{e}_{\rm b}$ and longitude of ascending node in the xy-plane , respectively.  These angles are calculated as 

\begin{equation}
    i_{\rm polar} = \arccos{\left( \frac{L_z}{L} \right)}
    \label{eq:inc_polar}
\end{equation}
and
\begin{equation}
    \AN_{\rm polar} = \arctan{\left( \frac{L_y}{L_x} \right)}.
    \label{eq:omega_polar}
\end{equation}

Precession of the disk causes $\AN_{\rm polar}$ to circulate.  As the disk oscillates, inclination damping causes the oscillation amplitude to shrink over time, allowing the disk to settle towards a polar orientation and $\theta_{\rm polar}$ to settle towards 0.
For fixed disk density and temperature distributions, the timescale for polar alignment $t_{\rm damp}$ is inversely proportional to $\alpha$ \citep{King2013,Lubow2018}, and so disks with a higher $\alpha$-viscosity settle towards a polar alignment quicker, performing less oscillations before reaching a polar configuration.

From Figure \ref{fig:angle}, the polar alignment timescale can be estimated to be roughly $300$ binary orbits for the $\alpha = 0.1$ simulation, and roughly $500$ binary orbits for the $\alpha = 0.05$ simulation.  This is roughly consistent with the analytical predictions given by Equations 29-30 of \cite{Lubow2018}, which give $t_{\rm damp} = 220\ T_b$ and $455\ T_b$, respectively.  The difference in density evolution of these two cases likely accounts in part for how $t_{\rm damp}$ departs from a pure $1/\alpha$ dependence as $\alpha$ changes.  For simulations with viscosities $\alpha < 0.05$, the oscillation amplitude is not reduced significantly during the course of our simulation time, suggesting that the timescale for polar alignment is at least on the order of several thousand binary orbits.  The polar alignment timescales for the $\alpha = 0.01$ simulation are also similar to the SPH simulations with high resolution ($10^6$ particles).  SPH simulations at lower resolution show a quicker dampening of the tilt oscillations and thus a lower alignment timescale.

We measure the precession periods in our $\alpha = 0.1$ and $\alpha = 0.01$ simulations to be roughly $300-400$ binary orbits.  The precession period increases over time as the disk material is redistributed outwards.  These values are in rough agreement with the SPH simulations of the same Shakura-Sunyaev viscosity\footnote{The equivalent SPH artificial viscosity for these simulations is $\alpha_{\rm{SPH}} = 4$ and $0.4$ respectively \citep{Lodato2010}.} in \cite{Martin2017} and \cite{Martin2018} (see their Eq. 2), which measure precession periods of roughly $300$ binary orbits for disks that start with a tilt of $80^\circ$.

The settling behavior towards polar alignment is shown clearly in Figure \ref{fig:phase}, which plots the trajectory of the disk in the $i \, \mathrm{\cos\AN} - i \, \mathrm{sin\AN}$ phase space, along with trajectories of test particles in gray.  After an initial damping phase, present in all simulations, the disk trajectories spiral inwards towards the point $(0.0, 0.5)$, corresponding to a disk oriented at exactly $90\degree$ inclination.  High-viscosity disks $(\alpha \gtrsim 0.05)$ spiral inwards and align towards polar orientation quickly.  Disks with lower viscosities $(\alpha < 0.05)$ follow nearly closed oscillating trajectories close to the $70\degree$ test particle trajectory, with lower values of $\alpha$ settling towards trajectories slightly closer to a polar orientation.  As noted above, the SPH simulations that start at an inclination of $80^{\circ}$ with $\alpha=0.01$ show a significant decay of tilt over 1000 binary orbital periods.




\begin{figure}
    \centering
    \includegraphics[width=\columnwidth]{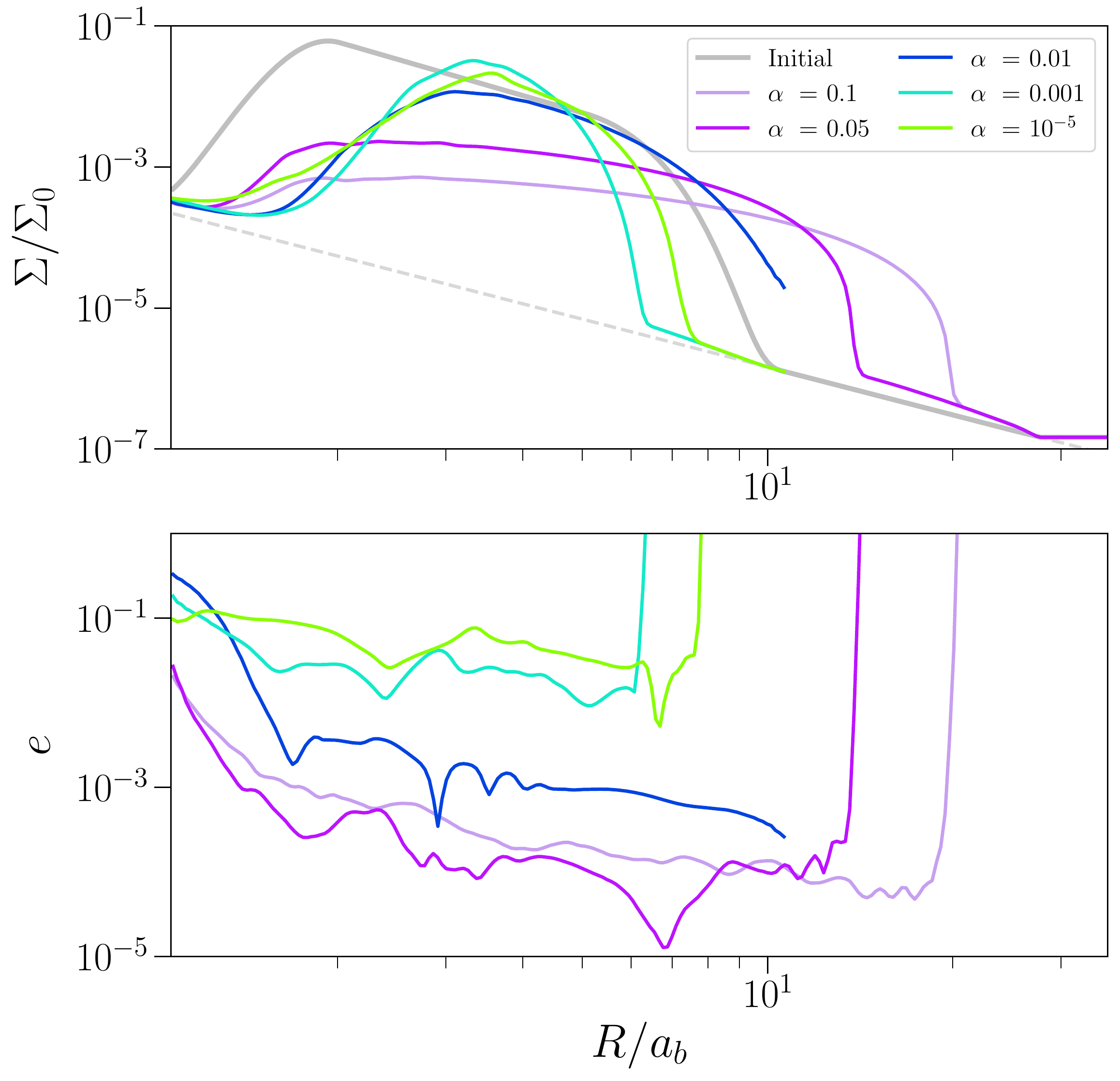}
    \caption{\emph{Top:} Spherically integrated radial surface density profiles at $t = 1000$ binary orbits, plotted on a logarithmic scale.  \emph{Bottom:}  Radially averaged disk eccentricity at $t = 1000$ binary orbits, calculated with Equation 16 of \protect\cite{Shi2012}.  }
    \label{fig:sd}
\end{figure}

\begin{figure*}
    \centering
    \includegraphics[width=0.95\textwidth]{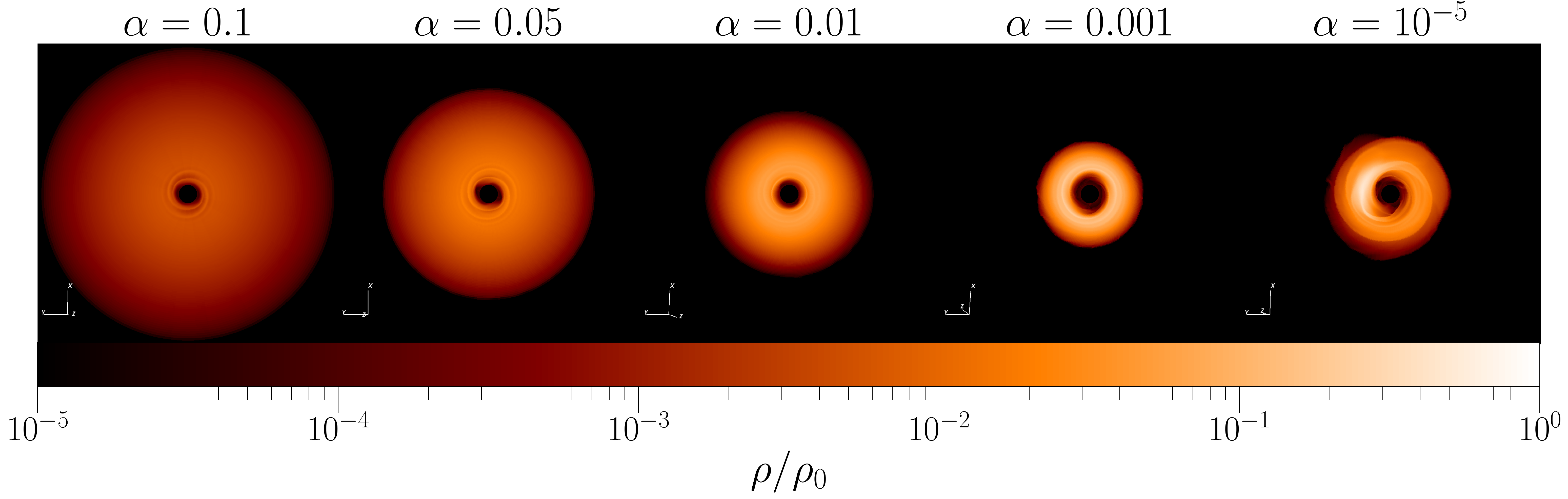}
    \caption{Face-on density profiles of the disk at $t=675$ binary orbits.}
    \label{fig:faceon}
\end{figure*}

\begin{figure}
    \centering
    \includegraphics[width=\columnwidth]{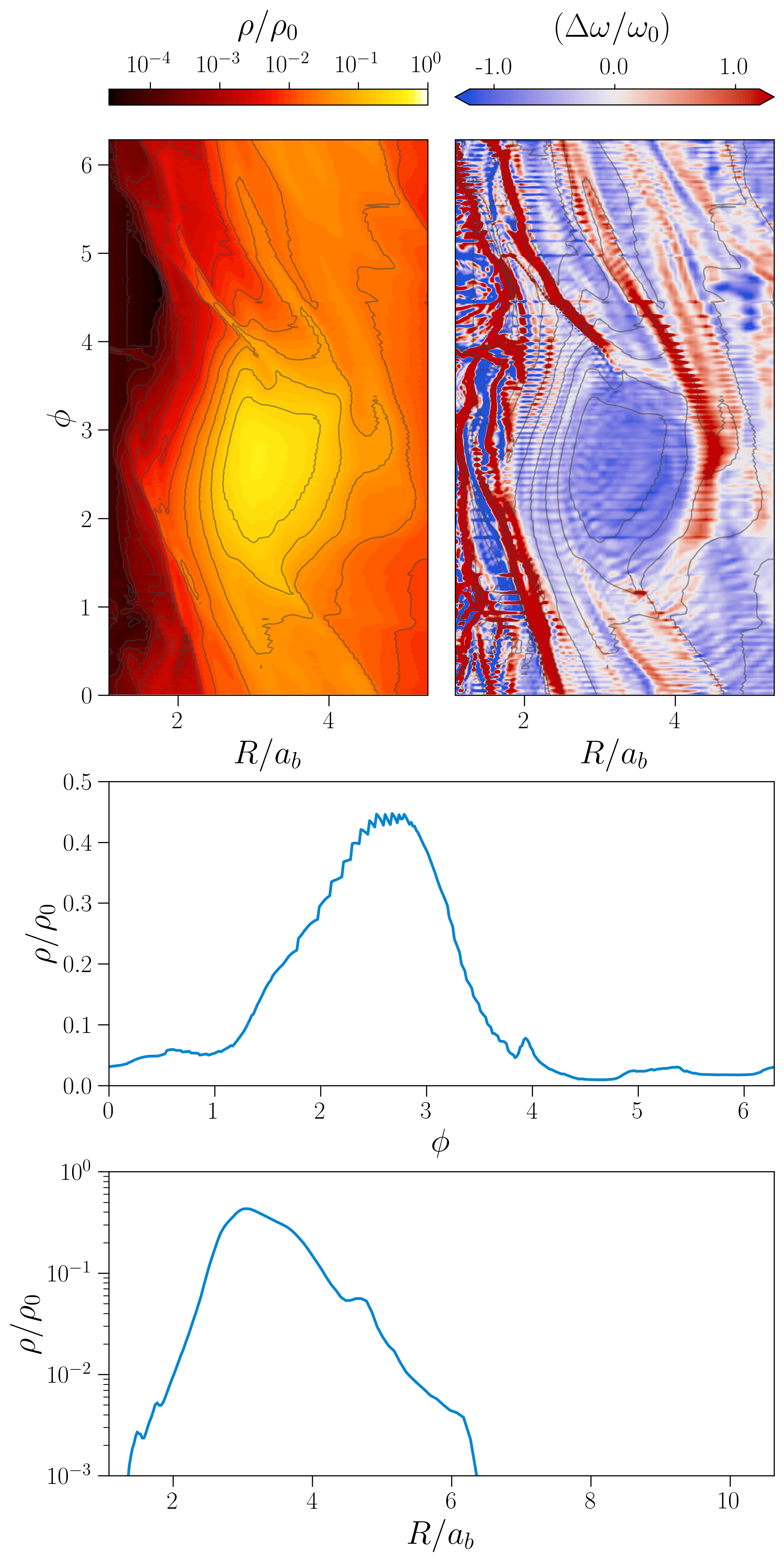}
    \caption{Closeup of the vortex in the $\alpha = 10^{-5}$ simulation at $t=692$ binary orbits. \emph{Top left:} Gas midplane density.  The local overdensity and spiral arm pairs are clearly visible.  \emph{Top right:} Local vorticity.  Strong anticyclonic regions are indicated in blue.  Spiral arms are visible as red diagonal lines in the vicinity of the vortex.  In both plots, black curves show contours of density.  \emph{Middle:} Azimuthal density across the vortex center. \emph{Bottom:} Radial density across the vortex center. }
    \label{fig:vortexcloseup}
\end{figure}




Figure \ref{fig:sd} shows the disk surface density profiles at $t=1000$ binary orbits.  The inner edge of the disk drops off quickly in all simulations at roughly $2a$, with higher viscosity disks exhibiting a smaller inner cavity.  
This behavior is in agreement with previous numerical simulations \citep{Franchini2019}, as well as analytical predictions \citep{Miranda2015, Lubow2018} suggesting that disks in a polar orientation should be truncated somewhat closer to the binary than in the coplanar case. The model of \cite{Miranda2015} suggests that the disks in these simulations should be should be truncated at the 1:3 commensurability for an outer Lindblad resonance. 
Streams of gas that flow into the central gap may carry much or all of the gas required for a steady-state accretion disk, as is known in the coplanar case \cite[e.g.,][]{Artymowicz1996, Shi2012, Munoz2016}. The profiles generally follow a wider, thinner distribution with increasing $\alpha$-viscosity.  Since the diffusion coefficient $\nu$ is proportional to $\alpha$, the disks with a larger $\alpha$ are able to spread out further due to their shorter viscous timescale $t_{\rm visc} \sim R^2/\nu$.  As the disk material spreads radially, the change in mass distribution increases the disk's moment of inertia, increasing the precession timescale as seen in Figure \ref{fig:angle}.
Notably, the case of $\alpha = 10^{-5}$ does not follow this trend, having a slightly wider surface density profile than the $\alpha = 10^{-3}$ simulation. This may be due to the formation of vortices at low values of $\alpha$, which generates spiral wakes in the disk.  These spirals can increase the outward transport of material and widen the surface density profile.  We discuss the details of vortex formation and their impacts on the disk in Section \ref{sec:vortex}.



The bottom panel of Figure \ref{fig:sd} shows the shell-averaged eccentricity of the disk at $t = 1000 $ binary orbits.  We calculate the eccentricity of the disk using Equation 16 of \cite{Shi2012}.  The disk eccentricity increases and eventually saturates as time increases, so these values can be taken as a maximum value for each simulation.  We find that simulations with $\alpha \gtrsim 0.01$ are able to maintain low eccentricities, while disks with lower viscosities are excited to moderate $(e \sim 0.1)$ eccentricity.

\begin{figure*}
    \centering
    \includegraphics[width=0.95\textwidth]{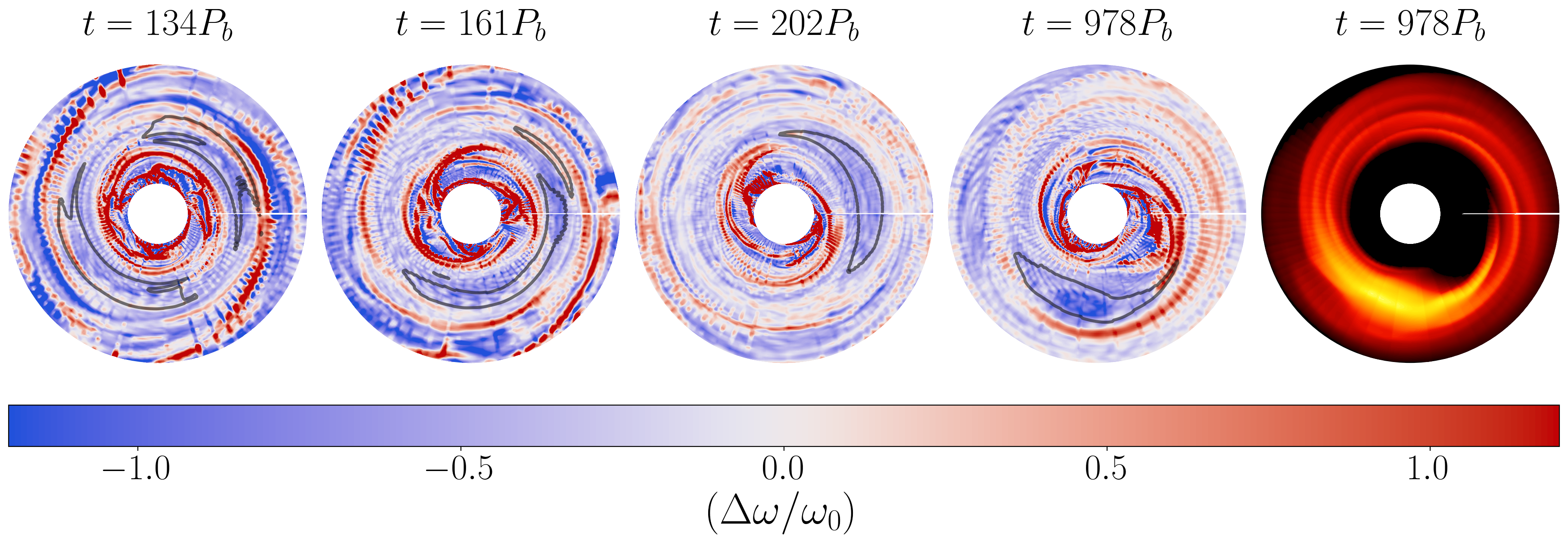}
    \caption{Snapshots of the $\alpha = 10^{-5}$ simulation, showing the local disk vorticity. In the left four panels, blue regions denote areas of large anticylonic vorticity.  The black contours highlight high-density regions $(\rho = 0.6 \rho_{\rm max})$ within the disk midplane.  \emph{Far Left:} Early on, the RWI triggers with an $m = 2$ mode, creating two vortices on opposite sides of the disk.  \emph{Left:} Differences in orbital velocities cause the vortices to move together and merge.  \emph{Middle:} The result is a single vortex which orbits at the inner edge of the disk.  \emph{Right:} The remaining vortex is long-lived, and persists until the end of the simulation.  \emph{Far Right:} Midplane disk density, showing the vortex as well as the multiple spiral arms.  The density scale is shifted compared to Figure \protect\ref{fig:vortexcloseup} in order to highlight the spiral arms.}
    \label{fig:vortex}
\end{figure*}

\subsection{Vortex Formation}
\label{sec:vortex}

\begin{figure}
    \centering
    \includegraphics[width=\columnwidth]{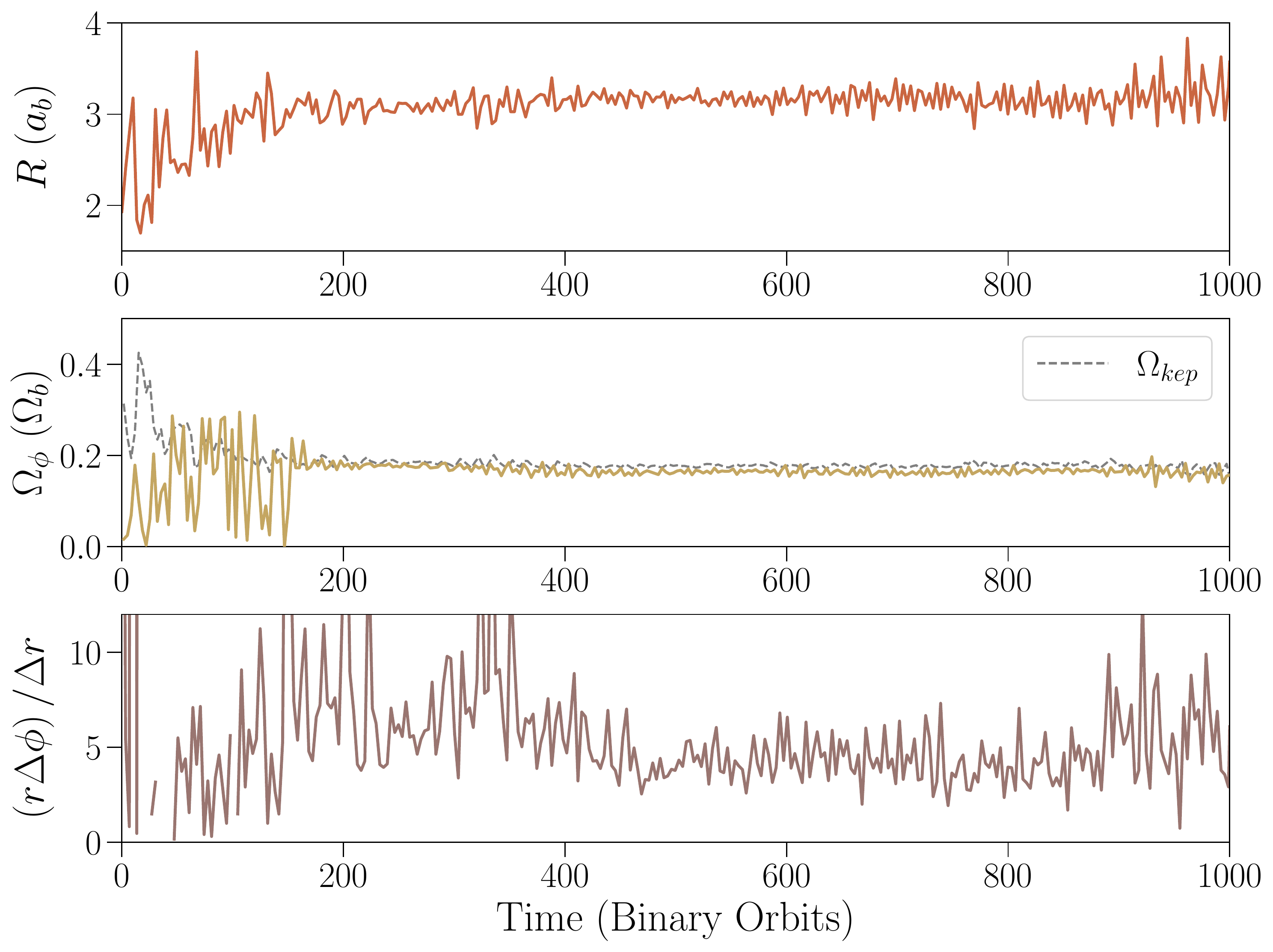}
    \caption{Orbital motion of the vortex in the $\alpha = 10^{-5}$ simulation.  \emph{Top:} Orbital radius of the vortex. \emph{Middle:} Orbital frequency, in units of the binary orbital frequency $\Omega_b$.  The dotted line shows the Keplerian orbital speed at the vortex's orbital radius.  \emph{Bottom:} Vortex aspect ratio $\chi = r \Delta \phi / \Delta r$.  }
    \label{fig:vortexspd}
\end{figure}

A gallery of disk midplane density profiles is shown in Figure \ref{fig:faceon}.  Each panel shows a face-on view of the disk at $t = 675$ binary orbits.  Similar to Figure \ref{fig:sd}, material in disks with higher viscosity is more spread out with a lower surface density.  Some light spiral features are present towards the inner edge of the disk in all simulations, which vary in strength as the disk precesses and changes its orientation relative to the binary, but most of the disks are azimuthally symmetric and largely featureless.

The notable exception to this behavior is the nearly inviscid $\alpha = 10^{-5}$ simulation, which shows a prominent density enhancement along the inner edge of the disk, as well as an associated one-armed spiral.  Horseshoe-shaped overdensity features have previously been observed in ALMA dust continuum images \citep{Casassus2013,vanderMarel2013,Caassus2016}, and have been explained by disk material moving on eccentric orbits \citep{Ragusa2017}, or vortices generated by sharp surface density gradients via the Rossby Wave Instability (RWI, \citealt{Lovelace1999,Li2000}).  Our simulations are the first instances where vortices have been observed within polar disks.  The steep surface density profile along the inner edge of the disk creates a minimum in the local vortensity, an important characteristic in generating RWI vortices \citep{Bae2015}.  Vortex formation can be suppressed by small amounts of viscosity $(\alpha \lesssim 10^{-3})$ \citep{ValBorro2007,Fu2014,Zhu2014b,Owen2017}, so they were not observed in previous SPH simulations of polar disks.

A close-up of the overdensity feature is shown in Figure \ref{fig:vortexcloseup}.  We calculate the midplane density and vorticity by transforming the disk data as outlined in Appendix \ref{sec:appendix}.  The left panel shows the disk density, where the overdense region is clearly shown.  The right panel shows the local disk vorticity $(\omega-\omega_0)/\omega_0$, where  $\omega = -(\nabla \times \mathrm{v}_\theta)$ and $\omega_0 = \frac{1}{2} \Omega_K$ is the initial Keplerian vorticity.  At the disk midplane after the coordinate transformation, the unit vector in $\theta$ points downwards, opposite in direction to the unit vector in $z$, so we add a negative sign to our calculation of vorticity to match the sign convention of vorticity in other coordinate systems, i.e. $\omega = \nabla \times \mathrm{v}_z$.  Large areas of anticyclonic motion are coincident with the overdensity region, and confirm that the overdense region is associated with the creation of a vortex.  The density cuts across the vortex show that the peak gas density $\rho_{max}$ is enhanced by a factor of $\gtrsim 8$ times the background, which is consistent with previous simulations of RWI vortices (e.g. \citealt{Lyra2008,Bae2015}).

Two spiral arms are visible on the leading side of the overdensity, extending along slanted density contours and trailing inwards towards the star.  A similar pair of outward pointing spirals is present along the trailing edge of the vortex, but are less defined, causing the two spirals to partially combine and form a single, wide spiral arm behind the vortex.  In the vorticity plot, the spirals are visible in red as diagonal lines of high positive vorticity.  Together, these create two distinct pairs of spiral arms originating from the vortex:  a ``central'' pair of spirals, which begin along the center line of the vortex (roughly $3\ a_{\rm b}$ in the figure) and extend directly away from the vortex, and a ``tangential'' pair of arms, which are created along the inner and outer radial edges of the vortex and run tangent to the edge of the vortex.  



The initial triggering of the RWI and creation of the vortex follows the evolution outlined in \cite{Bae2015}.  We show panels of the vortex growth and evolution in Figure~\ref{fig:vortex}.  The large vortensity minimum at the inner edge of the disk generates several RWI vortices during the linear growth phase of the instability (far left).  The number of initial vortices can vary, but our simulations show a consistent $m = 2$ mode initially present within the first tens of binary orbits.  The initial appearance of this mode may be related to the binary components, as previous single-star numerical calculations find the $m = 3-5$ modes to have the fastest growth rates \citep{Li2000,Lin2012}.  Once the vortices are formed, differences in their orbital speeds will cause them to migrate towards each other and merge (left), eventually forming a singe anticyclonic vortex.  The merging process is relatively quick, and all vortices are combined into a single vortex within a few hundred orbits of their creation (middle).  The remaining vortex persists until the end of the simulation (right).  A comparison with the disk density (far right) shows the coincident overdensity and spiral arm features as in Figure \ref{fig:vortexcloseup}.

The orbital motion of the vortex is shown in Figure \ref{fig:vortexspd}.  We locate cells of density $\rho > 0.5\rho_{max}$ within the vortex and calculate the average of their locations as the vortex radius and azimuthal position in the local disk coordinates.  We also use the radial and azimuthal extent of these cells to calculate the aspect ratio of the vortex $\chi = r \Delta \phi / \Delta r$.  After the initial vortices merge, the resulting single vortex orbits at roughly 3 times the binary separation at Keplerian speed.  The vortex aspect ratio is somewhat more variable even after the vortices have finished merging, but is generally around a value of $\chi \sim 5$.



\section{Discussion}
\label{sec:discussion}

\subsection{The effect of vortices vs. lumps in the disk}
The overdensity feature observed in the $\alpha = 10^{-5}$ simulation is noticeably different than similar features created in coplanar circumbinary disks.  Several coplanar disks have been found to exhibit horseshoe features, but these features have been explained as a density ``lump'', created by eccentric gas at the inner edge of the disk interacting with inflow streams created by the action of the binary torque \citep{Shi2012,Miranda2017,Ragusa2017,Ragusa2020}.  In contrast, the overdensity feature in the polar disk corresponds to the vorticity minimum in the disk, and is closer in nature to vortex-induced clumps caused at the edges of gaps opened by massive planets \citep{ValBorro2007,Zhu2014,Hammer2017}.

Overdensities in circumbinary disks can drive periodic accretion onto the central binary.  Simulations of coplanar curcumbinary disks and the binary components by  \cite{Munoz2016} show the lump at the inner edge causes pulsed accretion at the rate of the lump's orbital period.  When binary eccentricity is included, the accretion flows become pulsed at the rate of the binary orbital period, and can preferentially favor one star for several hundred orbits.  In \cite{Miranda2017}, mass accretion is found to vary around circular binaries on short timescales of $\frac{1}{2} T_b$, as well as on longer timescales of roughly $5 T_b$.

Accretion rates in polar disks have been less studied.  \cite{Smallwood2022} perform simulations of the HD 98800 system and find that binary accretion rates show no periodicity and are roughly constant in time for a polar disk with $\alpha = 0.01$.  The high viscosity in these simulations means they do not form vortices, even during the passage of the outer AaAb binary.  

\begin{figure}
    \centering
    \includegraphics[width=\columnwidth]{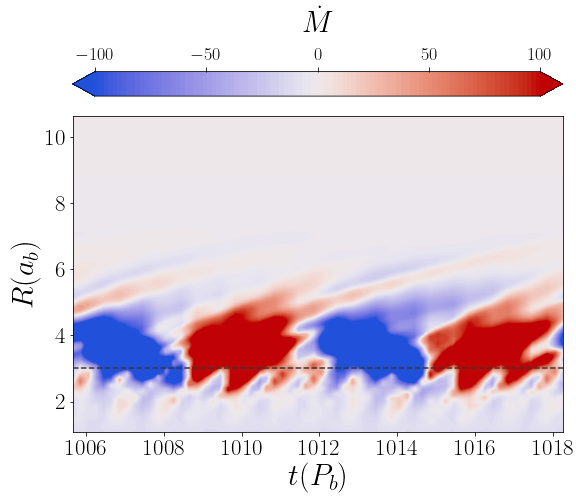}
    \caption{Shell-averaged radial mass flux of the $\alpha = 10^{-5}$ simulation.  The dashed line marks the approximate orbital distance of the vortex.  Short-period outflows are visible close to the binary $(2 a_b)$ as thin red stripes.  Long-period variability, associated with the orbital period of the vortex, is visible from $3-4 a_b$ as alternating red and blue bands.}
    \label{fig:vortexaccretion}
\end{figure}

The vortex in our $\alpha = 10^{-5}$ simulation exhibits similar variable behavior on both short and long timescales.  Figure \ref{fig:vortexaccretion} shows the shell-averaged radial mass flux over roughly 10 binary orbits.  Long-term flows are present at distances of $3-4 a_b$, oscillating with a period of roughly $5 T_b$.  This low frequency mass flux is due to the orbital motion of the vortex around the binary, carrying the overdense region on an eccentric orbit at its local orbital period (Figures \ref{fig:sd}, \ref{fig:vortexspd}).  Closer to the binary, at roughly $2 a_b$, short-term outflows are visible with a periodicity of roughly $T_b$.  These outflows are caused by small streams of material pulled from the inner edge of the disk and flung outwards by the binary torque, similar in nature to those seen by \cite{Shi2012}.  Since our inner binary is eccentric, these streams do not pile up and form an overdense lump \citep{Miranda2017}. Overall the vortex in our polar circumbinary disks is different from the lump in coplanar circumbinary disks in two ways: 1) the vortex only forms when $\alpha$ is very low, while the lump can form in disks with a much higher $\alpha$ (even in MHD simulations with $\alpha\sim0.1$, \citealt{Shi2012}); 2) the vortex exists in polar circumbinary disks with an eccentric binary, while the lump is only found when the binary is nearly circular ($e_{\rm b}\lesssim 0.05$, \citealt{Miranda2017}).

The overdensity created by the vortex changes the local pressure gradient on the neighboring gas, producing regions of super-Keplerian and sub-Keplerian velocity just inside and outside of the vortex orbital radius, respectively \citep{Kuznetsova2022}.  These deviations from Keplerian rotation may be observable as a kinematic signature \citep{Boehler2021}, which are inverted but similar in strength to the signatures formed in the gaps of protoplanetary disks thought to be formed by young planets \citep{Teague2018,Teague2019,Izquierdo2021}.  Future high-resolution studies with ALMA may be able to distinguish vortex creation in polar disks.

Observations of HD 98800 B from \cite{Kennedy2019} show a uniform disk in both gas and dust, with no evidence of large scale disk asymmetry.  The absence of vortex-like features may be a sign that the viscosity in HD 98800 B is high enough to prevent vortices from forming.  Another explanation could be that vortices are relatively transient features within the disk.  The lifetime of vortices in the disk is on the order of thousands of orbital periods \citep{Zhu2014b,Hammer2021}.  For HD 98800 B, this corresponds to a lifetime of $\lesssim 1000$ years for a vortex formed at the inner edge of the disk, which is much shorter than the disk lifetime of $\sim 10$ Myr \citep{Barrado2006}.  Therefore, it is possible that any vortices that initially formed in the disk have long since dispersed, unless vortices are continuously generated by the inner binary.

The rightmost panel of Figure \ref{fig:vortex} shows the vortex and spiral arms as they appear in the disk.  Notably, the appearance of the spiral arms resembles the scattered light images of HD 142527 \citep{Avenhaus2014,Avenhaus2017}, which is also known to have an overdensity feature \citep{Casassus2013}.  Spirals along the inner edge of the disk can be created, without the creation of a vortex, due to the gravitational interaction of the binary with the disk \citep{Price2018}, so it is unclear if vortex-generated spirals are the cause of the spiral arms in HD 142527.

\subsection{Prospects for Polar Circumbinary Planets}
It is unknown if circumbinary planets are able to form within polar disks.  Currently, no planets have been observed in polar orbits around binaries, nor planet candidates in polar-aligned disks.  The lack of detections is at least partially due to a strong observational bias; current methods for detecting exoplanets rely on identifying data with multiple, periodic transits, which favor single-star, coplanar systems.  Planets on circumbinary orbits exhibit large transit timing variations, on the order of hours or days \citep{Armstrong2013}.  Inclined orbits are more likely to produce single or irregular transits due to the orbital precession induced by the binary torque \citep{MartinD2014,Chen2019,Chen2022}, rendering most detection methods used for single planets unusable and requiring the use of special transit folding methods (e.g. \citealt{MartinD2021}).  If circumbinary planets are equally likely to form in polar disks as they are in coplanar disks, they may be as common as planets around single-star systems \citep{Armstrong2014,MartinD2014}.


The formation of vortices within low-viscosity disks can have significant effects on planet formation history.  An outstanding problem in planet formation involves the formation of large planetary embryos in a disk, as gas drag from the disk can cause growing particles to rapidly spiral inwards onto the star, before they are able to decouple from the gas \citep{Weidenschilling1977}.  Anticyclonic vortices in a disk act as grain traps, and can intercept over half of the grains that cross its orbit during their inward radial migration \citep{Fromang2005}.  Once captured in the vortex, the local vorticity works to focus particles towards the vortex ``eye'', which can facilitate accelerated growth of planetary material all the way up to Jupiter-mass planets \citep{Klahr2006,Lyra2008,Zhu2014,Owen2017}.  In polar disks, vortices may also lead to direct planet formation or seed the disk with a large initial planetesimal population, which can allow further planetary growth and formation terrestrial planets in polar orbits \citep{Childs2021,Childs2022}.  

Figures \ref{fig:vortexcloseup} and \ref{fig:vortexspd} show that the vortex orbits at a few times the binary separation, close to the inner edge of the disk.  Indeed, the sharp surface density profile created at the inner cavity wall allows the RWI to produce vortices in the absence of viscosity.  This is close to the theoretical stability limit for circumbinary planets \citep{Dvorak1986}.  Curiously, many of the circumbinary planets discovered by \emph{Kepler} and \emph{TESS} orbit just outside the inner stability limit \citep{MartinD2014,Welsh2014}.  Together, this may suggest a potential in-situ formation scenario for close-in circumbinary planets, where the sharp inner cavity of the circumbinary disk generates long-lived vortices which allows for efficient trapping of solid materials.  Rapid planet formation at this distance would produce circumbinary planets close to the circumbinary inner stability limit.


\section{Conclusion}
\label{sec:conclusion}

In this paper, we describe the first grid-based simulations on the evolution of polar-aligning circumbinary disks. We find the disks align towards a polar orientation, with timescales in rough agreement with previous SPH studies and analytic estimates.  In nearly inviscid disks, RWI vortices can be generated close to the disk inner edge while polar alignment occurs.  These vortices can persist for thousands of binary orbits in a polar disk, and create overdensities within the disk.  Two pairs of spiral arms are seen originating from the vortex, along the center and tangent to the vortex core.  The overdensities created by the vortices are fundamentally different from similar features created through eccentricity excitation within the disk.  The combination of vortices and spiral arms may create a detectable signature in polar disks, though none have been observed as of yet.  The presence of vortices and overdensities in polar circumbinary disks may also aid in the formation of polar circumbinary planets, accelerating planet growth close to the inner stability limit.

\section*{Acknowledgements}
This material is based upon work supported in part by UNLV's Faculty Top Tier Doctoral Graduate Research Assistantship Grant Program (TTDGRA) and the National Aeronautics and Space Administration under Grant No. 80NSSC20M0043 issued through the Nevada NASA Space Grant Consortium. Z. Z. acknowledges support from the National Science
Foundation under CAREER Grant Number AST-1753168. RGM and SHL acknowledge support from NASA through grants 80NSSC21K0395 and 80NSSC19K0443. Figures in this paper were made with the help of Matplotlib \citep{Hunter2007}, NumPy \citep{Harris2020}, and VisIt \citep{HPV:VisIt}.  The animations included in the online version of this paper were made with the help of the NAS Visualization Team.

\section*{Data Availability}
The data used in this paper is available upon request to the corresponding author.

\bibliographystyle{mnras}
\bibliography{ref}


\appendix

\section{Coordinate Transform of Disk Data}
\label{sec:appendix}

Simulations of disks are usually oriented so that their local coordinate axes are aligned with the simulation axes.  This alignment simplifies the analysis of important disk parameters such as azimuthal slicing along the midplane, or calculation of vorticity along the disk's vertical axis.

Disks that are inclined, warped, or precess over time are not always aligned with the simulation axes, and so these quantities are harder to acquire without some prior manipulation.  We show here our method of manipulating the simulation data of an inclined disk so that it may be analyzed in a similar way as a flat, zero-inclination disk.  As in the paper, we use spherical-polar coordinates for this section, though the basic process can be used for any coordinate system.

Consider a disk that is inclined to the simulation axes at an angle $i$.  Any point in 3D space can be described with \emph{simulation coordinates} $(r, \theta, \phi)$ or \emph{disk coordinates} $(r, \theta', \phi')$, which are not necessarily the same.  The transformation between these two coordinate systems is given by the equations
\begin{equation}
    \sin^2{\theta'} = \left( \sin{\theta} \cos{\phi} \right)^2 + \left( \sin{\theta} \sin{\phi} \sin{i} + \cos{\theta} \sin{i} \right)^2
    \label{eq:coord1}
\end{equation}
and
\begin{equation}
    \tan{\phi'} = \frac{ \sin{\theta}\sin{\phi} \cos{i} + \cos{\theta} \sin{i} }{ \sin{\theta} \cos{\phi} }.
    \label{eq:coord2}
\end{equation}

Equation \ref{eq:coord1} is equivalent to Equation 44 of \cite{Zhu2019}.  Since this equation only gives $\sin^2{\theta'}$, a degeneracy exists when trying to map coordinates on the top and bottom halves of the sphere.  Therefore, we calculate $\cos{\theta'} = \sqrt{1-\sin^2{\theta'}}$ and multiply the value by $-1$ if the original polar angle $\theta$ lies below the equator of the mapped coordinates.  This corresponds to 
\begin{equation}
    \theta = \frac{\pi}{2} - \tan^{-1}{\left( \tan{i}\sin{\phi} \right)}.
\end{equation}
To construct an untilted disk, we start by creating an evenly spaced grid along the disk coordinates $\theta'$ and $\phi'$.  For each grid cell, we transform the grid coordinates to simulation coordinates, locate the nearest data cell in the original simulation coordinates, and copy that cell data from the simulation into the grid cell.  The result is a data cube oriented along the disk axes, which can be analyzed using traditional methods.

For velocity, additional transformations must be done to account for the rotation of the velocity components.  The azimuthal distance element in the disk coordinates $d\phi'$ can be expressed in terms of the simulation coordinates as
\begin{equation}
    d\phi'^2 = (rd\theta)^2 + (r\sin{\theta}d\phi)^2.
\end{equation}
This allows us to calculate the angle between the disk and simulation grids $\psi$ with
\begin{equation}
    \tan{\psi} = \frac{d\theta}{\sin{\theta}d\phi}.
\end{equation}




In practice, we calculate $\psi$ by comparing adjacent cells in the disk coordinates and converting back to the simulation coordinates, allowing the distances to be calculated using simple differences.  Once the local rotation angle is found, the velocity in the disk coordinates $\vel_\theta'$ and $\vel_\phi'$ can be calculated with
\begin{equation}
    \vel_\theta' = \vel_\theta \cos{\psi} - \vel_\phi \sin{\psi}
\end{equation}
and
\begin{equation}
    \vel_\phi' = \vel_\theta \sin{\psi} + \vel_\phi \cos{\psi}.
\end{equation}
The radial velocity remains unchanged during the rotation, i.e. $\vel_r' = \vel_r$.  

\begin{figure}
    \centering
    \includegraphics[width=\columnwidth]{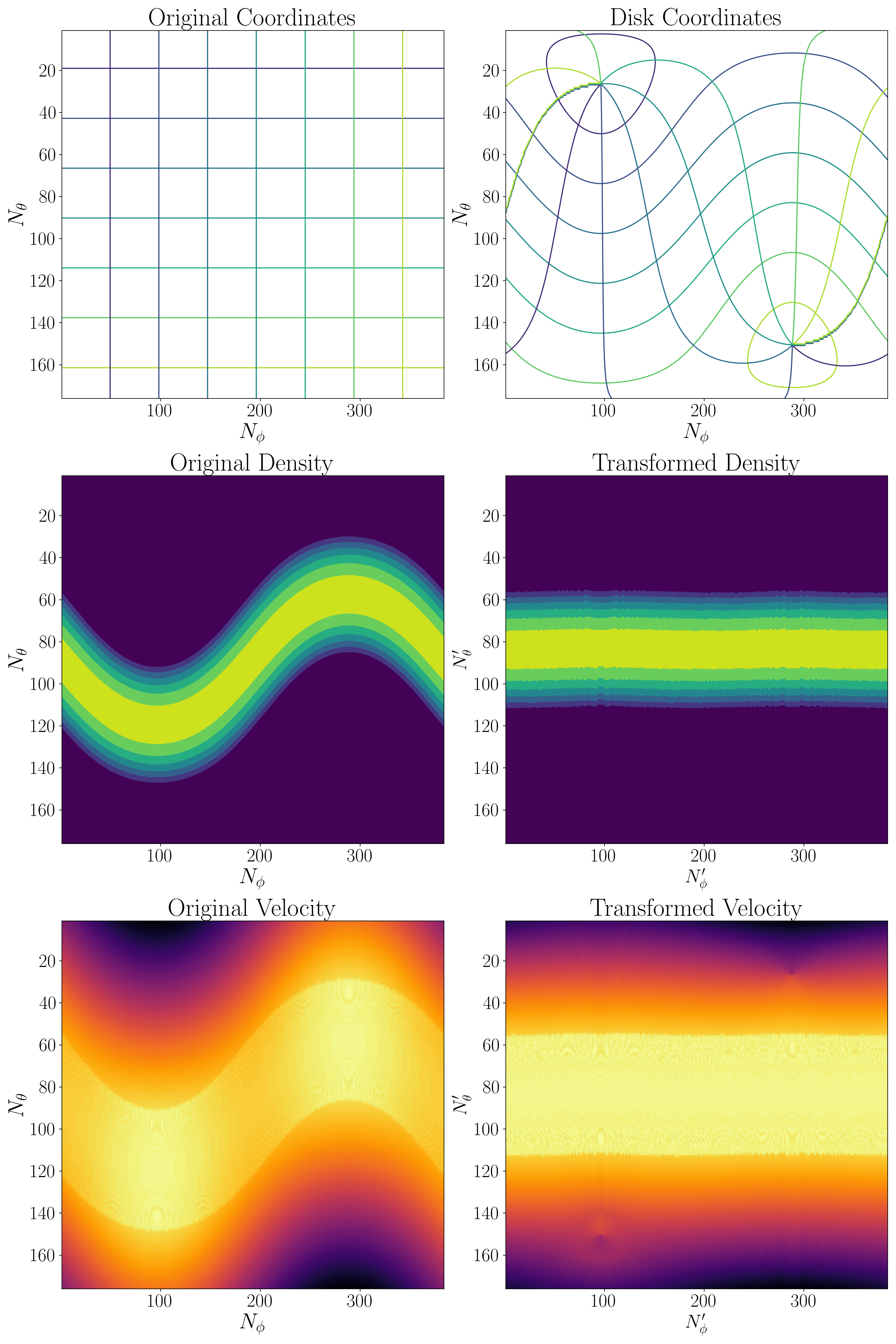}
    \caption{Example of the coordinate transform using a 2D slice of the simulation data at $t=0$.  \emph{Top Row:} Coordinate grid lines for the simulation and disk coordinates. \emph{Middle Row:} Disk density in the simulation and disk coordinates.  \emph{Bottom Row:} Azimuthal velocity $\vel_\phi$ in the simulation and disk coordinates.}
    \label{fig:transform}
\end{figure}

Figure \ref{fig:transform} shows an example of the transformation applied to a 2D spherical slice of our simulation data at $t=0$.  The transformed density and velocity are similar to that of a disk aligned with the coordinate system.  Small artifacts are visible in the areas close to the poles of the simulation coordinates due to the lack of initial data in these regions.  Small vertical striations can also be seen along the transformed data, 

For simulations in this paper, we transform the disk data using the total angular momentum vector $\vhat{\mathbf{L}}$ to calculate $i$ and $\AN$ using Equations \ref{eq:inc} and \ref{eq:omega}.  Warped or broken disks can be analyzed in a similar fashion by using the local angular momentum vector $\vhat{\mathbf{L}}(R)$ to transform each annulus separately.

\bsp	
\label{lastpage}
\end{document}